\documentclass[final,english]{bullsrsl}

\usepackage[latin1]{inputenc}
\usepackage[T1]{fontenc}

\usepackage{natbib} 
\usepackage{graphicx}
\usepackage{graphicx}

\begin{document}
\title{From Sky to Science: Time Domain Astronomy with the ILMT Survey}

\author[affil={1}, corresponding]{Kuntal}{Misra}
\author[affil={2,1}]{Bhavya}{Ailawadhi}
\author[affil={3,4}]{Talat}{Akhunov}
\author[affil={1,5}]{Monalisa}{Dubey}
\author[affil={1,5}]{Naveen}{Dukiya}
\author[affil={6}]{Sara}{Filali}
\author[affil={7,8}]{Paul}{Hickson}
\author[affil={1}]{Priyanshi}{Kumari}
\author[affil={1}]{Gokul Singh}{Mehra}
\author[affil={9}]{Vibhore}{Negi}
\author[affil={1,10}]{Kumar}{Pranshu}
\author[affil={6}]{Jean}{Surdej}
\author[affil={1,10}]{Sarvesh Kumar}{Yadav}

\affiliation[1]{Aryabhatta Research Institute of Observational Sciences (ARIES), Manora Peak, Nainital-263002, India}

\affiliation[2]{Physical Research Laboratory, Navrangpura, Ahmedabad, Gujarat-380009, India}

\affiliation[3]{National University of Uzbekistan, Department of Astronomy and Astrophysics, 100174 Tashkent, Uzbekistan}

\affiliation[4]{Ulugh Beg Astronomical Institute of the Uzbek Academy of Sciences, Astronomicheskaya 33, 100052 Tashkent, Uzbekistan}

\affiliation[5]{Mahatma Jyotiba Phule Rohilkhand University, Pilibhit Bypass Road, Bareilly 243006, Uttar Pradesh, India
}
\affiliation[6]{Institute of Astrophysics and Geophysics, Li\`ege University, All\'ee du 6 Ao\^ut 19c, 4000 Li\`ege, Belgium
}
\affiliation[7]{Department of Physics and Astronomy, The University of British Columbia, 6224 Agricultural Road, Vancouver, BC V6T 1Z1, Canada
}

\affiliation[8]{Outer Space Institute, The University of British Columbia, 325-6224 Agricultural Road, Vancouver, BC V6T 1Z1, Canada
}

\affiliation[9]{Kavli Institute for Astronomy and Astrophysics, Peking University, Beijing 100871, China}

\affiliation[10]{Department of Applied Optics and Photonics, University of Calcutta, Kolkata, 700106, India}

\correspondance{kuntal@aries.res.in}

\date{}
\maketitle

\begin{abstract}
The PyLMT transient detection pipeline has been operational since November 2023 and has detected $\sim$10,000 transients and variable objects in ILMT images in near real time. The survey has enabled the discovery and characterisation of a range of bright and faint optical transients. The detected population spans a broad diversity of astrophysical phenomena, including supernovae, variable stars, active galactic nuclei (AGN) variability, and Solar System objects such as asteroids and comets. We provide a concise overview of the detection of a variety of transient and variable classes identified with the ILMT, emphasising the survey's potential contributions to a broad class of astrophysical sources. We highlight representative science cases, including supernovae, AGNs, variable stars, and asteroids, to demonstrate the scientific potential of the ILMT survey.
\end{abstract}


\section{Introduction}
\label{introduction}
The 4-m International Liquid Mirror Telescope \citep[ILMT;][]{2025A&A...694A..80S} is currently the only liquid mirror telescope in the world dedicated to astronomical observations. Operating in a zenith-pointing, time-delayed integration (TDI) mode, the ILMT repeatedly monitors a fixed strip of the sky at a cadence of one day, making it the first optical time-domain survey facility in India. The telescope has a field of view of approximately $22'$, and conducts observations in the SDSS \textit{g}$'$, \textit{r}$'$, and \textit{i}$'$ filters and covers $\sim$36 deg$^{2}$ of sky area per night with a median seeing of $\sim 1''.2$. With a limiting magnitude of $\sim$22 in the \textit{g}$'$ filter \citep{2024BSRSL..93..820A}, the ILMT is sensitive to both bright and faint optical transients, enabling systematic studies of time-variable phenomena. 

To support the transient search programme, a dedicated automated pipeline, \texttt{PyLMT} \citep{2025MNRAS.538..133P}, was developed for transient detection and classification of transient alerts. The pipeline employs image subtraction techniques for candidate identification, followed by a convolutional neural network \citep[CNN;][]{726791}-- based classification to distinguish transient candidates according to their host galaxy morphology. The classification scheme includes \textit{extended-host}, \textit{point-source}, and \textit{hostless} categories, facilitating rapid classification of candidates for further inspection. Using this framework, the ILMT transient search programme has led to the detection and independent discovery of a diverse range of astrophysical phenomena, including supernova candidates, variable stars, variability in active galactic nuclei (AGN), and solar system objects such as asteroids and comets. In the era of other large surveys like the ZTF, ATLAS, Pan-STARRS, and LSST, the ILMT occupies a niche in terms of cost-effectiveness, a uniform 1-day cadence of observations of the zenith sky, and faint limiting magnitude owing to an aperture size of 4 m and minimal extinction of the zenith sky.

This paper is organised as follows: Section~\ref{PyLMT} presents a technical overview of the \texttt{PyLMT} transient detection pipeline. Section~\ref{Detections} discusses representative examples from the detected transient and variable population. The key points of this work and future prospects are summarised in Section~\ref{summary}.

\section{Transient Search Framework}
\label{PyLMT}

The \texttt{PyLMT} pipeline was developed to enable automated, reliable transient searches within the ILMT data stream. Its primary objective is to identify transient and variable sources in newly acquired science images while suppressing artefacts introduced during image subtraction. At its core, \texttt{PyLMT} employs the image subtraction module \texttt{ILMTDiff}, which isolates time-variable sources by removing the static and non-varying sources from science images. For each new observation, a reference image acquired at the same Local Sidereal Time (LST) is selected. The science and reference images are aligned using World Coordinate System (WCS) information, followed by flux scaling and 2D convolution-based point-spread function (PSF) matching to account for variations in observing conditions. Subtraction of the convolved reference from the science image yields a difference image consisting of both the transient and artefacts/bogus sources. Candidate sources detected in the difference images are evaluated using a CNN-based real/bogus classifier implemented in the \texttt{TransiSearch} module, producing a high-purity sample of astrophysical candidates. Confirmed detections are subsequently classified using the \texttt{NovaNet} module based on host morphology into \textit{hostless}, \textit{point-source}, and \textit{extended-host} categories. The accuracy of the real/bogus classifier ranges from 94~$\%$ to 98~$\%$ while that for the candidate classifier is 98.6 $\%$. The overall detection precision was 0.91. The detection recall rate was determined to be more than 0.9 for candidates brighter than 21 mag and around 0.5 for candidates fainter than 21.5 mag. Catalogue cross-matching is integrated at this stage to reject known solar system objects and identify previously catalogued sources using external databases such as that of Institut de m\'ecanique c\'eleste et de calcul des \'eph\'em\'erides
\citep[IMCCE;][]{2006ASPC..351..367B}, \textit{Gaia} \citep{2016A&A...595A...2G}, and \texttt{SIMBAD} \citep{2000A&AS..143....9W}.

Together, these components form an end-to-end automated framework that enables systematic detection and initial characterisation of transient and variable sources in the ILMT data.

\section{Representative Science Cases}
\label{Detections}

This section discusses representative examples drawn from the transient and variable source population detected between the third (October 2023 -- May 2024) and the fourth (October 2024 -- May 2025) observation cycles. These cases are selected to illustrate the diversity of observed phenomenology, spanning different timescales, luminosities, and physical origins.

\subsection{Supernova candidates}

\begin{figure}
    \centering
    \includegraphics[width=\textwidth]{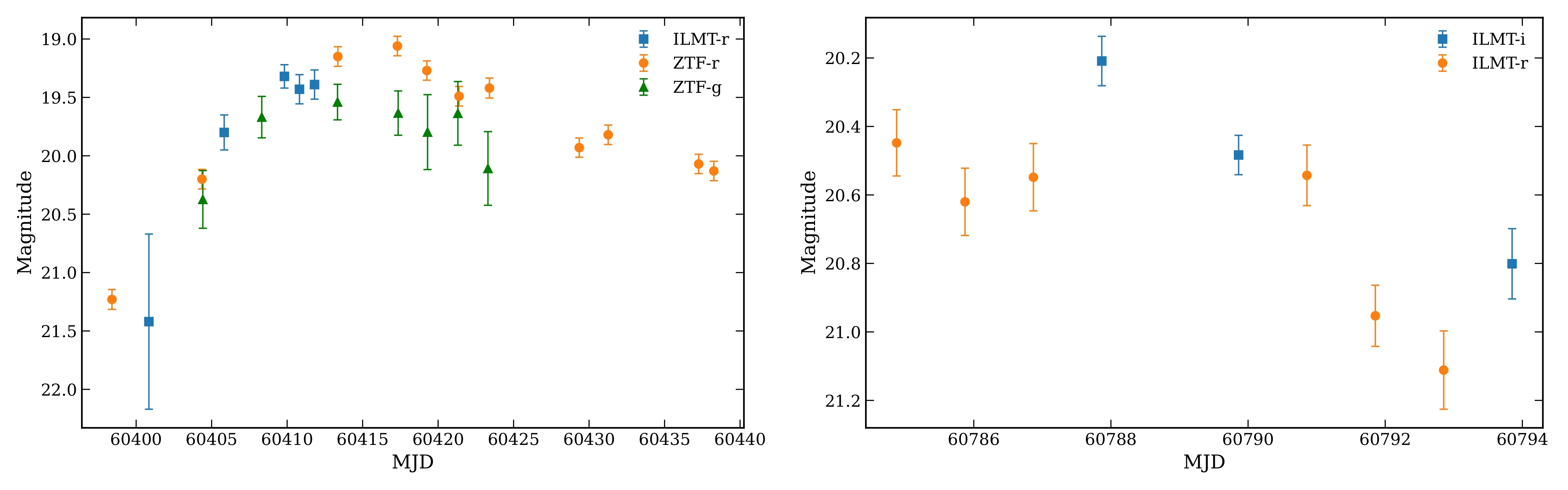}
    \vspace{-10pt}
    \begin{minipage}{12cm}
    \caption{Multiband light curves of supernova candidates AT~2024fxn (left) and AT~2024aifv (right). ZTF data points are also included in the light curve of AT~2024fxn.}
    \label{fig:SNe}
    \end{minipage}
\end{figure}

Supernovae are extragalactic transient phenomena associated with either the core collapse of massive stars or the thermonuclear disruption of white dwarfs. These events typically occur within host galaxies and are therefore classified as \textit{extended-host} candidates by \texttt{PyLMT}. As part of the transient search programme with the ILMT, a total of 22 supernova candidates were identified, including six independent discoveries. On average, an SN candidate is detected once every month. Figure~\ref{fig:SNe} presents the optical lightcurves of two newly discovered supernova candidates, AT~2024fxn \citep{2024TNSTR.964....1P} and AT~2024aifv \citep{2025TNSTR1468....1P}. Most of the detected candidates were faint, having discovery magnitudes between 19--21. This prevented any spectroscopic follow-up observation, and hence, definitive spectral classification could not be performed for these events. More details on supernova candidates are presented in \citet{2025BINA...Dubey}.

\subsection{Variable Stars}

\begin{figure}
    \centering
    \includegraphics[width=\textwidth]{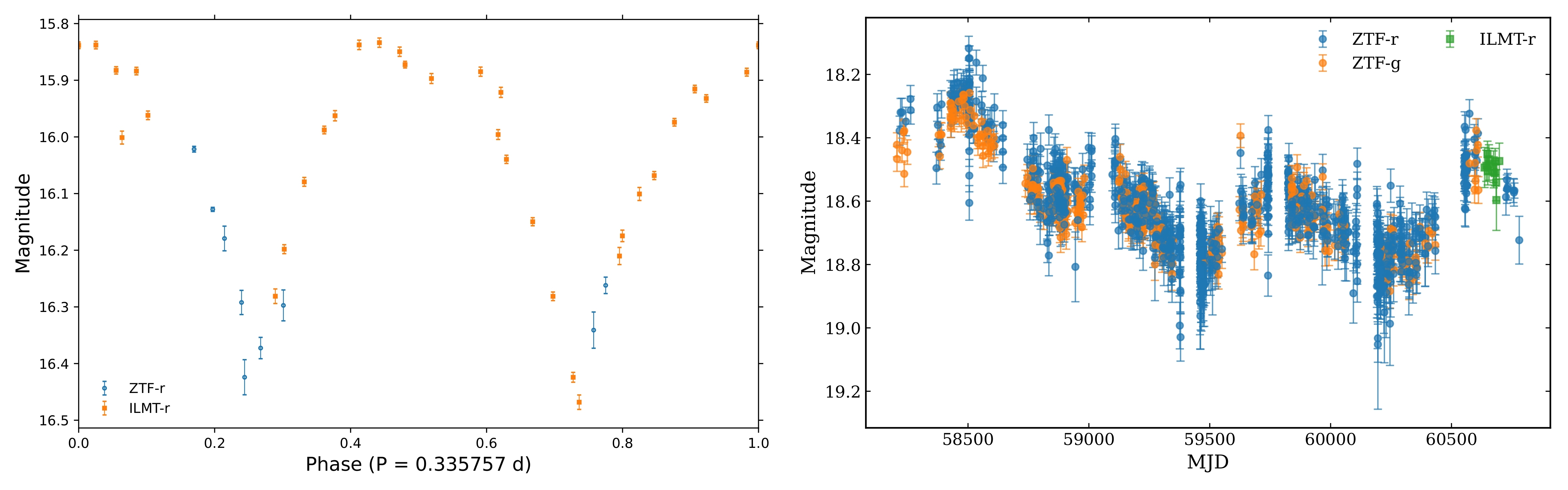}
    \vspace{-10pt}
    \begin{minipage}{12cm}
    \caption{Left: Phase-folded light curve of the ILMT-detected eclipsing binary J152019.4+292928. Right: ILMT light curve of the quasar ILMT083233.2+292039 supplemented with ZTF data.}
    \label{fig:AGN_VS}
    \end{minipage}
\end{figure}

Depending on the observing season, the ILMT field of view intersects regions near the Galactic plane, yielding a rich harvest of stellar variability detections. This has enabled the detection of a diverse population of variable stars, including eclipsing binaries, RR~Lyrae stars, T~Tauri objects, $\delta$~Scuti variables, and rotating variable stars. In addition to periodic variability, several eruptive and outburst events were identified in cataclysmic variable systems, highlighting the utility of the ILMT survey to both short- and long-timescale stellar variability. These sources are primarily identified as \textit{point-source} candidates by the \texttt{PyLMT} pipeline, consistent with their stellar nature. Variability in more than 500 variable stars has been detected with the pipeline till May 2025, including the identification of 15 new variable stars. The regular cadence and homogeneous coverage provided by the ILMT allow well-sampled lightcurves to be constructed for such objects, facilitating reliable identification of variability patterns. The lightcurve of one representative variable star detected by the ILMT is shown in Figure~\ref{fig:AGN_VS} (left).

\subsection{Active galactic nuclei}

AGNs are the luminous central regions of some galaxies, powered by accretion onto supermassive black holes. Depending on their spectral properties and orientation, AGNs are classified into subclasses such as quasars, Seyfert galaxies, and blazars. A defining characteristic of AGNs is their intrinsic variability, which can span a wide range of amplitudes and timescales. A significant fraction of the variable sources detected by the ILMT are classified as quasars, which are typically identified as \textit{point-source} candidates in the \texttt{PyLMT} pipeline. Because image subtraction is used, the transient search in ILMT is naturally biased toward detecting AGNs with relatively large-amplitude variability. This characteristic makes the ILMT dataset particularly well-suited for the selective identification of AGNs displaying strong optical variability, which could also be candidates for exotic cases such as changing-look AGNs and other AGNs exhibiting extreme variability. Variability in nearly 130 AGNs has been detected with the pipeline. Figure~\ref{fig:AGN_VS} (right) shows the lightcurve of one of the quasars detected with the ILMT. Automated detection and photometric monitoring of quasars with the ILMT are presented in \citet{2025BINA...Akhunov}.

\subsection{Asteroids}

The most numerous class of detections during the transient search is asteroids. These objects are typically classified as \textit{hostless} candidates by the pipeline, as they are not associated with any persistent astrophysical host in the reference images. Over the first four observational cycles, approximately 9000 distinct and previously catalogued asteroids have been detected in ILMT data. This high detection yield is primarily enabled by ILMT's faint limiting magnitude.  The detected asteroid candidates are cross-matched against publicly available solar system object ephemeris catalogues, including those maintained by the Minor Planet Center (MPC) and the IMCCE. Owing to the fixed zenith-pointing and TDI observing mode of the ILMT, most asteroids are detected only a limited number of times as they traverse the field of view, except in cases where their apparent motion is aligned with the detector readout direction, allowing repeated detections. Given the high astrometric precision of the ILMT, typically better than $0''.1$ \citep{2026PASP..138g4506N}, the positional measurements of these asteroids provide valuable constraints for refining their orbital parameters. The detection and identification of asteroids with the ILMT are presented in \citet{2025BINA...Filali}.

\section{Summary and Future Prospects}
\label{summary}

The ILMT has been conducting a systematic zenith-sky survey in three spectral bands. To enable an automated search for transient and variable sources, the dedicated \texttt{PyLMT} pipeline was developed and deployed. This effort has resulted in the detection of nearly 10,000 transient and variable objects, including 22 supernovae candidates, more than 500 variable stars, nearly 130 AGNs, nearly 9000 asteroids, and a few comets, over a period from October 2023 to May 2025. 

Beyond the individual detections, this work demonstrates that a liquid mirror telescope, despite its fixed pointing and unconventional design, can be effectively utilised for scientific-grade time-domain astronomy. The successful implementation of an end-to-end transient search framework with the ILMT highlights the potential of liquid mirror facilities to contribute meaningfully to systematic time domain studies. Continued monitoring with the ILMT, complemented by archival data from surveys such as ZTF, will enable more detailed characterisation of the detected sources, with more comprehensive results to be presented in future works (Pranshu et al. 2026; under review).

\begin{acknowledgments}
\small{
We thank the referee for providing constructive comments on the manuscript, which improved the clarity of the presentation. The 4-m International Liquid Mirror Telescope (ILMT) project results from a collaboration between the Institute of Astrophysics and Geophysics (University of Li\`{e}ge, Belgium), the Universities of British Columbia, Laval, Montreal, Toronto, Victoria, and York University, and Aryabhatta Research Institute of Observational Sciences (ARIES, India). The team acknowledges the contributions of ARIES's technical and scientific staff. JS wishes to thank Service Public Wallonie, F.R.S.-FNRS (Belgium), and the University of Li\`{e}ge, Belgium, for funding the construction of the ILMT. PH acknowledges financial support from the Natural Sciences and Engineering Research Council of Canada, RGPIN-2019-04369. PH, SF, and JS thank ARIES for hospitality during their visits to Devasthal. We sincerely acknowledge Anna Pospiezalska for her significant contributions to ILMT functioning. MD acknowledges the Innovation in Science Pursuit for Inspired Research (INSPIRE) fellowship award (DST/INSPIRE Fellowship/2020/IF200251) for this work. KP acknowledges the support from the Erasmus+ Programme of the European Union for a research visit to the Institute of Astrophysics and Geophysics, University of Li\`ege, Belgium. TA thanks the Ministry of Higher Education, Science, and Innovations of Uzbekistan (grant FZ-20200929344). JS and KM acknowledge the assistance received from the ANRF SERB-726 VAJRA Faculty Scheme, India. KM acknowledges the support from the BRICS grant DST/ICD/BRICS/Call-5/CoNMuTraMO/2023 (G) funded by the DST, India. We acknowledge the Belgo-Indian Network for Astronomy and astrophysics (BINA) consortium approved by the International Division, Department of Science and Technology (DST, Govt. of India; DST/INT/BELG/P-09/2017) and the Belgian Federal Science Policy Office (BELSPO, Govt. of Belgium; BL/33/IN12), for allowing us to present our work in the form of a paper in the proceedings of the fourth BINA workshop.}

\end{acknowledgments}

\begin{furtherinformation}

\begin{orcids}
\orcid{0000-0003-1637-267X}{Kuntal}{Misra}
\end{orcids}

\begin{authorcontributions}
Toward the completion of this work, all authors have made significant contributions.
\end{authorcontributions}

\begin{conflictsofinterest}
The authors declare no conflict of interest.
\end{conflictsofinterest}

\end{furtherinformation}

\bibliographystyle{bullsrsl-en.bst}

\bibliography{BINA_kuntal}

\end{document}